# X-Ray spectroscopy and timing (XSPECT) experiment on XPoSat - instrument configuration and science prospects

**Radhakrishna V,[a*] Anurag Tyagi,[a] Koushal Vadodariya,[a] Vivek K Agrawal,[a] Rwitika Chatterjee,[a] Ramadevi M C,[a] Kiran M Jayasurya,[a] Kumar,[a] Vaishali S,[a] Srikar P Tadepalli,[a] Sreedatta Reddy K,[a] Lokesh K Garg,[a] Nidhi Sharma,[a] Evangelin L Justin,[a]**

[a]U R Rao Satellite Centre, Indian Space Research Organisation, Bengaluru, INDIA

**Abstract**: X-ray Polarimeter Satellite (XPoSat) with POLarimeter Instrument in X-rays (POLIX), is India's first spacecraft dedicated to study medium energy X-ray polarisation from celestial objects. X-Ray Spectroscopy and Timing (XSPECT) instrument on XPoSat is configured to study long term spectral behaviour of select sources in Soft X-ray regime. The instrument uses Swept Charge Devices (SCD)s to provide large area and spectral performance with passive cooling arrangement. The instrument consists of set of collimators with two different FOVs, optical light blocking filters, and signal processing electronics. The instrument was designed, tested and calibrated on ground. The unique opportunity is provided by ISRO's XPoSat mission, where a source is observed for longer duration. The device used also enables spectroscopy study of brighter sources compared to the CCD based spectrometers. The first results demonstrate instrument capability for spectral studies in the 0.8 keV-15 keV energy band.

**\*** Radhakrishna V, E-mail: rkrish@ursc.gov.in

## 1    Introduction

XPoSat is the India's first X-ray polarimetric mission launched on 1st January 2024 by PSLV C58 rocket from Sriharikota, India. The satellite carries two co-aligned instruments, POLarimeter Instrument in X-rays (POLIX) and X-ray SPECtroscopy and Timing (XSPECT). POLIX will investigate the X-ray polarization in cosmic X-ray source in medium energy X-rays (8 - 30 keV) and XSPECT will carry out X-ray spectroscopy of these sources in soft X-ray (0.8 - 15 keV) band.

Astrophysical sources emitting predominantly in X-rays have been typically associated with binary systems. These systems, where a compact object such as a black-hole, a neutron star which is an end product of stellar evolution accretes matter from a normal main sequence star are called X-ray binaries. X-ray binaries are classified as low-mass or high mass X-ray binary based on the mass of the companion star. In low-mass X-ray binaries (LMXBs) accretion proceeds through Roche lobe overflow and the in-falling material loses its potential energy and angular momentum and forms an accretion disc around these stars, which emit X-rays [1,2]. In high-mass X-ray binaries (HMXBs) accretion proceeds via stellar wind. Neutron stars in HMXBs have high magnetic field (~ $10^{12}$ Gauss) and hence can direct the accreted matter on to its magnetic pole, producing beam of X-rays. Hence they show X-ray pulsations with period equal to the rotation period of neutron star [3].

XSPECT payload operates in the soft X-ray band of 0.8 – 15 keV with good spectroscopic resolution (< 200 eV at 6 keV) and moderate timing (2 ms) capability. In comparison to other spectroscopy missions (currently in orbit) effective area of XSPECT is smaller. But, SCDs used in XSPECT provide faster readout in comparison to CCDs. This has the advantage of high flux handling capability (> 2000 count/s without photon pileup) compared to CCDs.

Taking advantage of long duration (2-4 weeks) observation required by POLIX payload to measure the X-ray polarization, XSPECT payload will be able to carry out continuous and long term spectral and temporal studies of X-ray sources. Moreover, spectrosopic information derived from XSPECT will complement the polarization measurement from POLIX onboard XPoSat. Such observations are unique and first time. Some of the key science that XSPECT can address are; Understanding the nature of the soft excess [4] and studying variability and origin of soft excess in the X-ray pulsar, investigating the evolution of spin period, emission mechanisms and pulse profile, measuring the spin and mass of the blackholes through continuum and iron line profile fitting [5], studying evolution of X-ray spectrum in NS-LMXBs (neutron star low-mass X-ray binaries) and nature of soft thermal component [6], study of low frequency quasi-periodic oscillations (QPOs) in NS-LMXBs and blackhole X-ray binaries in soft X-ray band [7].

XSPECT payload is accommodated in India's XPoSat mission launched on 1st of January 2024 and PV phase observation and initial calibration has been successfully completed. In the first year, long-duration observations (2-4 weeks) of several bright X-ray sources have been carried out which include observations of black-hole X-ray binaries, X-ray pulsars, neutron star low-mass X-ray binaries (NS-LMXBs) and super-nova remnants (SNR). For removal of the observational systematics of POLIX payload the S/C spins around the source pointing axis at the rate of 0.2 RPM. This paper introduces XSPECT instrument with configuration details and ground performance.

## 2    Instrument Configuration

XSPECT science is around spectroscopic and timing capabilities of the instrument and does not require imaging capabilities. Large area detectors are chosen to meet the effective area requirement of the proposed investigations.

XSPECT is configured as three packages; two detector packages and one common electronics package. This configuration is worked out based on the accommodation possibilities on the spacecraft. The two detector packages are mounted on a bracket on either side of the POLIX instrument on the Payload Interface Module (PIM) deck of the spacecraft as shown in figure 1. The electronics package is mounted on one of the side deck. The detector packages are connected to electronics package through harness.

Each detector package has a detector sensitive/geometric area of 32 cm$^2$. Each detector package has field-of-view collimators, optical light blocking filters, detector modules, and amplifiers and the electronics package has back-end digital processing electronics, and interfaces to spacecraft systems, which are described in the following sections.

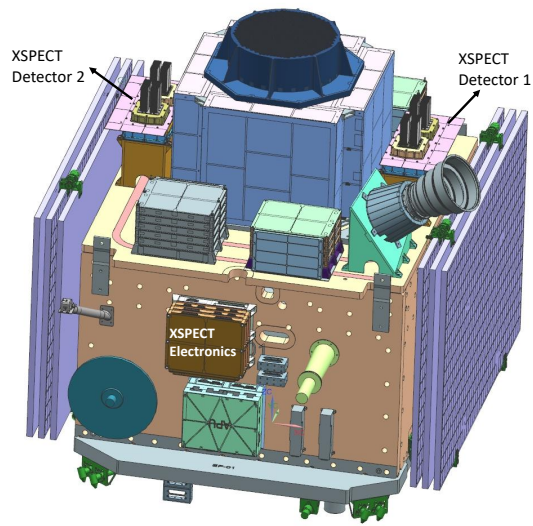

*Figure 1: XPoSat Spacecraft model with XSPECT mounted; two XSPECT detector modules shown are mounted on top deck and the electronics package is on one of the side deck of the spacecraft.*

*2.1 Detector system*

XSPECT instrument configuration is largely derived from Chandrayaan 2 Large Area Soft X-ray Spectrometer (CLASS), an X-ray fluorescence spectrometer, flown on Chandrayaan-2 [8,9]. XSPECT consists of an array of Swept Charge Devices (SCD) [10] operating in the energy range of 0.8 to 15 keV. SCDs are similar to X-ray CCDs except for the charge flow architecture on the device. Chandrayaan-1 [11] has proven SCDs as the best choice for the experiment where large collecting area is required along with good spectral resolution and high pile-up free count rate handling capacity. SCDs have no spatial resolution in the detector plane and hence are non-imaging devices. With a fast and continuous readout of SCDs, it yields good spectral resolution even at temperatures of -20°C. This avoids power demanding active cooling systems and permits the use of passive cooling systems like heat pipes and radiators to meet the thermal requirement.

The SCDs, CCD-236, are from Teledyne e2V and have also been flown on HXMT satellite for astronomical observations [12]. Major specifications of CCD236 are given in Table 1. The SCD is manufactured on epitaxial silicon wafers using the same processes as used for front illuminated CCDs. SiO2 and SiN are used as insulating layers underneath polysilicon electrodes. Each SCD has a total sensitive area of ~4 cm$^2$, internally divided into four sections. In each device, charge is transferred towards the centre of the device and merges from different sections, before transferring to a charge detection amplifier. Four devices are arranged as a quad as shown in figure 2. Total four quads, i.e., a total of 16 SCDs, form the full sensitive array of XSPECT instrument. Each SCD measures the energy of every detected photon which is stored and transmitted to the ground to generate X-ray spectrum. The time tag of each valid sample is the charge readout time corresponding to the respective event.

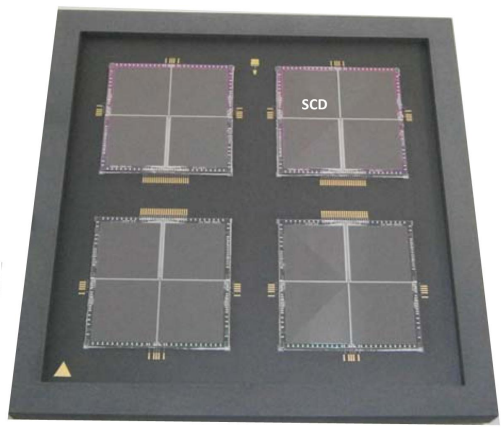

*Figure 2: A quad module with four SCDs on a ceramic package. Four such modules (i.e., 16 SCDs) are used in XSPECT.*

*Table 1: Major Specifications of CCD-236*

| Active area of unit SCD | ~ 4 cm2 |
|---|---|
| Device technology | epitaxial , scientific grade |
| Readout noise | < 10 electrons at 100 kHz |
| Readout rate | 10 – 100 kHz; optimized to operate at 100 kHz |
| Dark signal (@ -20° C) | < 10 electrons |
| Cycles to clear | 100 |
| Amplifier sensitivity | > 4 micro V/e |
| No. of clocking phases | Two-phase |
| Readout type | Continuous clocking |

### 2.2 Field Of View (FOV) Collimators

SCD is a non-imaging device i.e., it does not provide photon interaction pixel identification. We have adopted a passive collimator to restrict the FOV of XSPECT. As multiple bright sources may appear in the same FOV, it may be better to consider a smaller FOV to extract scientific information from the specific source. Contributions from the bright sources in FOVs of 1° X 1°, 2° X 2° and 3° X 3° have been looked into and the planned sources free of contamination from other sources in the field of view are 36, 27, and 19 respectively. While a smaller FOV is desirable, considering feasibility in hardware realization, square collimators of 2° X 2° and 3° X 3° FOV have been considered for XSPECT.

For 3° X 3° collimator, Al-6061 alloy is used and fabricated using wire Electrical Discharge Machining (EDM) method whereas for 2° X 2° collimator, additive manufacturing method is used with AlSi10Mg alloy powder. XRF characterisation was carried out to check for the contamination from the alloy material, especially around 6-7 keV where iron line at 6.46 keV from certain sources is of scientific interest. No significant line counts observed in the aforesaid energy range validating the materials use as collimator. Independent measurement of composition was also carried out by EDAX method. From the crab observation, the final estimated mean values of FOV are 1.94˚X 1.94˚ and 2.93˚ X 2.93˚ and the corresponding mean open area fractions are ~59% and ~75% respectively.

Out of the total 16 devices one device is shielded by 0.5 mm tantalum sheet, which will block X-rays in the dynamic range of interest and allow only GCR. Seven devices have 3 X 3 and eight devices have 2 X 2 FOV collimators. These three different set of devices together will help us better determine the 'near source' background as explained below;

All the collimators are co-aligned to spacecraft view axis which is aligned towards the sources of interest. On sky foot print of a 3 X 3 collimator completely engulfs the foot print of 2 X 2 collimator due to mutual alignment as shown in figure 3. X-ray sources of interest can be assumed to be point sources (angular size of source << FOV), hence both 2 X 2 and 3 X 3 devices will capture same source flux.

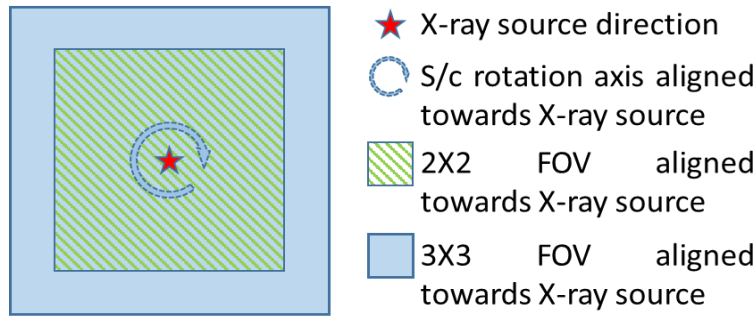

*Figure 3: Illustration of co-aligned FOVs pointing towards a source.*

Cosmic X-ray background (CXB) is be assumed to be uniform in the 3 X 3 FOV (also in the subset 2 X 2 FOV), hence will generate CXB background proportional to the solid angle that a collimator subtends on sky.

Due to assembly tolerances and dimensional errors, each collimator has their axes slightly misaligned with respect to spacecraft axis or source's direction. These were determined after instrument assembly on the spacecraft and verified on orbit by a source scan operation during the performance verification phase. These collimator axes 'offsets' will cause a flux loss by virtue of collimator angular response shown in the figure 4.

Since all the detectors are of the same physical area, galactic cosmic ray (GCR) events detected will be at same rate and hence, the x-ray blocked detector will provide a measure of only GCR associated background events.

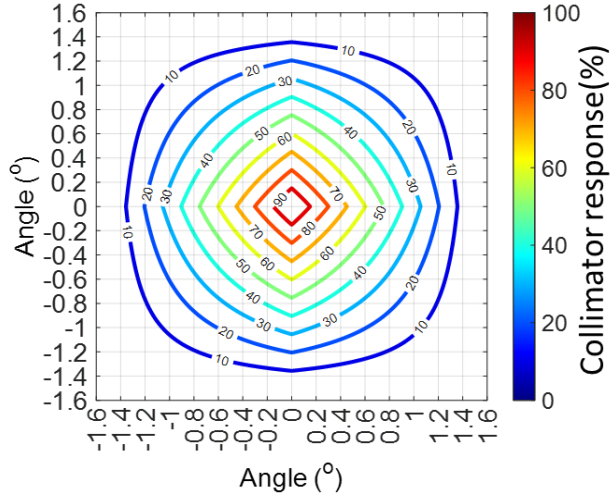

*Figure 4: Contours in the collimator response marks the ratio of flux reaching the detector to the flux entering the collimator opening (this is for 3 X 3 FOV).*

With the above design details, we can frame 3 equations as follows:

$F_{obs,2} = (S_0 \times A_2 \times OAF_2 \times collresp_2) \times DE + (0.25 \times B_0 \times A_2 \times OAF_2 \times FOV_2) \times DE + (P_0 \times A_2)$ (1)

$F_{obs,3} = (S_0 \times A_3 \times OAF_3 \times collresp_3) \times DE + (0.25 \times B_0 \times A_3 \times OAF_3 \times FOV_3) \times DE + (P_0 \times A_3)$ (2)

$F_{obs,15} = P_0 \times A_{15}$ (3)

Where,

| | |
|---|---|
| $F_{obs,2}$ | Total observed rate (counts/second) for 2X2 detectors (8 numbers) |
| $F_{obs,3}$ | Total observed rate (counts/second) for 3X3 detectors (7 numbers) |
| $F_{obs,15}$ | Total observed rate (counts/second) for blocked detector (1 number) |
| $S_0$ | Source flux in units of photons s$^{-1}$ cm$^{-2}$ |
| $B_0$ | CXB flux in units of photons s$^{-1}$ cm$^{-2}$ Sr$^{-1}$ |
| $P_0$ | GCR flux in units of events s$^{-1}$ cm$^{-2}$ (derived after modelling out the tantalum florescence lines in the blocked detector spectrum) |
| $DE$ | Detection Efficiency of the instrument |
| $OAF_2$, $OAF_3$ | Open area fraction of 2X2 collimators and 3X3 collimators respectively |
| $FOV_2$, $FOV_3$ | FOV solid angle in Steradian(Sr) of 2X2 collimators and 3X3 collimators respectively |
| $collresp_2$, $collresp_3$ | Average collimator response of 2X2 collimators and 3X3 collimators respectively |

| $A_2, A_3, A_{15}$ | Total geometrical area of 2X2 detectors, 3X3 detectors and blocked detector respectively. |
|---|---|

The factor of 0.25 in the second term of equations 1 and 2 is the average collimator response for the entire FOV as CXB uniformly extends across the entire FOV.

$S_0$, $B_0$ & $P_0$ can be determined by solving the above system of linear equations

## 2.3 Optical Blocking Filters (OBF)

SCDs are sensitive to visible light. The light blocking filter material and thickness chosen such that optical light transmission is less than $5 \times 10^{-6}$ but high X-ray transmission between 0.8 to 10 keV. A filter comprising 0.2 micron Al + 0.5 micron polyimide is used. The aluminium coating thickness is 0.1 micron each on the either side of the polyimide foil.

## 2.4 Effective area

Overall effective area, which is the sensitivity indicator determined by considering the detector geometric area, OBF X-ray transmission, detection efficiency, and the collimator open area fraction, is shown in figure 5. A correction factor due to the alignment of individual collimators with respect to the payload pointing axis is also considered.

The energy-dependent detection efficiencies of the SCDs and X-ray transmission efficiencies of OBF are calculated from the mass attenuation co-efficient provided by CXRO (https://henke.lbl.gov) in the given energy range.

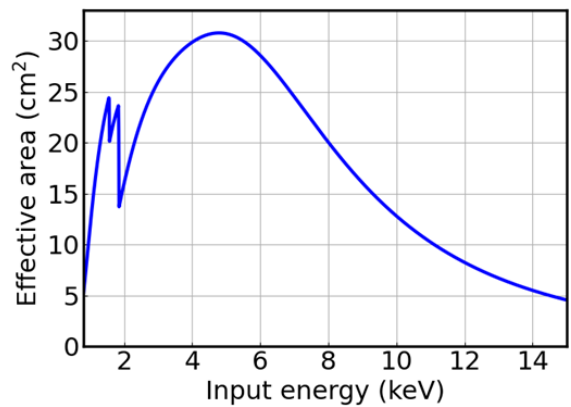

*Figure 5: XSPECT effective area considering open area fraction, detector efficiency and light blocking filter transmission (seven numbers of 3 x 3 collimators, eight numbers of 2 x 2 collimators)*

*2.5 Electronics configuration*

The XSPECT electronics has the following major elements shown in figure 6; (i) detector electronics, (ii) front-end electronics, (iii) digital processing electronics, (iv) spacecraft interfaces, and (v) power conditioning electronics.

Detector electronics is housed in detector package comprising rigid-flex-rigid card. Detector cards, have SCD quads (one per card) mounted on them. This card has the first stage of amplifier for each SCD and also contains passive components for filtering the bias voltages. This card is housed in detector package and is connected via a 44 pin ITT connector and harness to front end electronics card housed in electronics package through shielded RG316 cable. Front end electronics card consists of second stage of amplifiers and ADCs for signal digitization. AD8041, high bandwidth video amplifiers are followed by Correlated Double Sampling (CDS) based Analog Front End (AFE). These AFEs are provided with Single Event Latch up (SEL) mitigation circuit to protect from any latch up caused by particle event. The AFE digital data is sent to processing electronics for further processing.

Processing electronics card contains the RTAX2000S FPGA with all the logics built in for data processing. Each SCD quad module has a thermistor incorporated within the module for the measurement of temperature of the quad module. Temperature and SCD voltage monitoring circuit is incorporated in this card using a multi-channel 12-bits ADC. The temperature data is digitized every 256 ms.

Also, the high capacitive MOS driver-based clock generation circuits (phase clock and reset clock) are accommodated here. CCD-236 (SCDs) require 2-phase clocking and also two reset clocks. The clock frequency is optimised to 100 kHz. This temperature limit is set at -10 deg for XSPECT above which the clock is turned off.

The processing unit is configured to process the data of a Quad module. SCDs operate in continuous clocking mode where each sample is digitised and processed. When there is no event, the signal, i.e., 'zero energy' level, is estimated onboard for each SCD. An offset corresponding to desired LLD of 0.5 keV is added for setting the threshold above which a sample is considered an X-ray event. This offset is programmable and depends on the zero-energy level which is temperature dependent. Every 256 ms, 512 continuous samples are recorded for each detector and a histogram of this zero energy peak is created. The maximum count channel is considered as zero energy value used to compute the threshold value. Each X-ray event detected by any of the SCDs in a Quad is time-tagged and recorded in the data packet. The event data and the event time (14 bits) with the 2 bits of detector ID with 1 bit split event flag are stored for ~280 events and transmitted in one packet for each Quad. Each packet length is 4096 bytes. The packets are sent to Baseband Data Handling (BDH) system of the spacecraft through serial LVDS interface. XSPECT payload is programmed to have variable packet duration. Since the packet size is fixed, by varying the packet duration, data volume requirement in BDH is minimized. Packet duration, varies from 256 mSec to 16 sec, depends on the incoming X-ray flux.

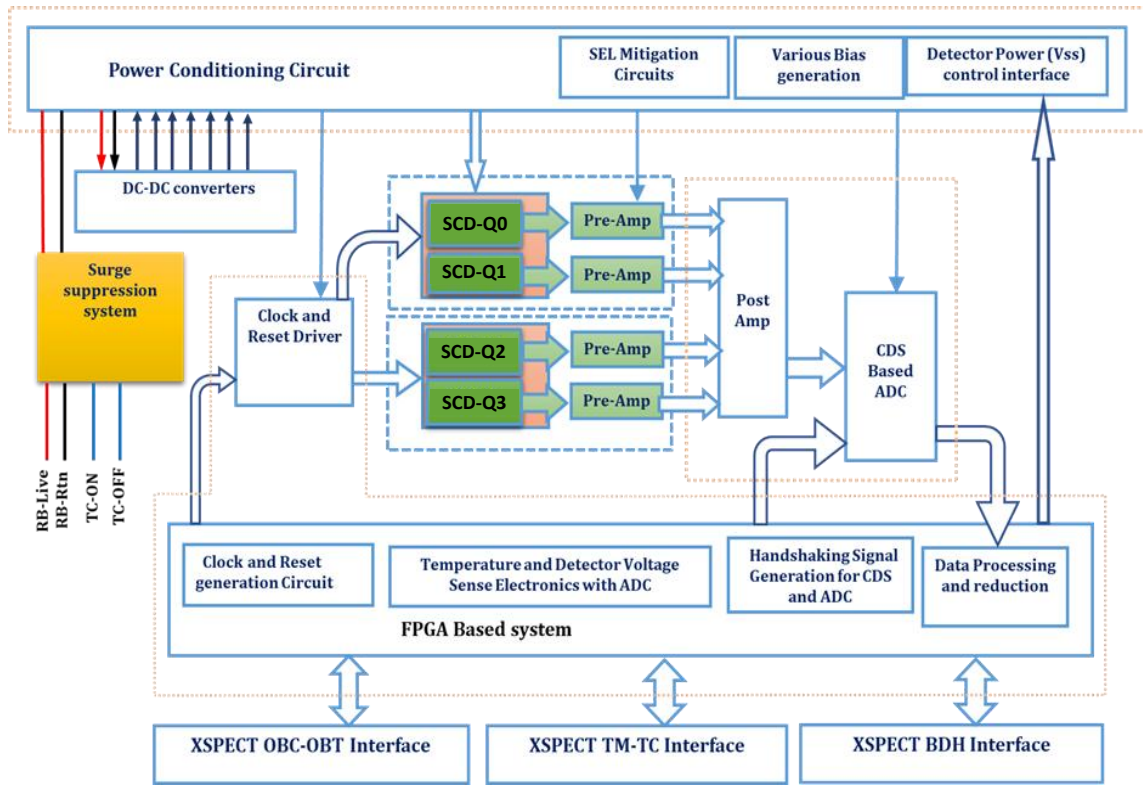

*Figure 6: Block diagram of XSPECT electronics elements.*

There is a dedicated serial interface with Spacecraft Onboard-Computer (OBC) for receiving the OBC time every second. Each event/X-ray interaction is logged along with time of arrival as recorded by instrument. This (arrival time) is a digital number and is converted to UT. Hence correlation of each photon's arrival time is known as UT.

Power conditioning electronics takes secondary voltages from DC-DC converters and all the SCD biases and other required voltages are generated here with LDOs and 3T voltage regulators.

This has relay and relay driver for XSPECT ON/OFF. XSPECT uses two numbers of DC-DC converters made in-house. There is provision to change the Substrate voltage of SCDs through ground command. The substrate voltage can be changed within a specified range of 9 V-11 V for mitigating the possible performance degradation by the radiation damage during mission life.

*2.6. Mechanical Configuration*

XSPECT is configured with 3 packages; two identical detector packages and one electronics package as shown in figure 1. Exploded view of a single detector package is shown in figure 7. Each quad module is mounted on aluminum base plate/radiator through a detector hold down clamp designed to hold the detector assembly and to facilitate expansions due to thermal load. The clamp is designed to accommodate thermal stress generated due to mismatch in CTE of materials and the launch vibration loads. Collimators are mounted on collimator base made of aluminium which also holds optical light blocking filters from inside. The detector pins pass through the slot provided on the base plate and the detector electronics card, a rigid – flexi - rigid PCB, is soldered and housed in detector electronics housing. The base/radiator plate is separated from the electronics housing by means of thermal spacers to provide thermal insulation. The realised detector package is shown in figure 8(a).

Both the detector packages are connected to a common electronic package through harness. The electronic package, shown in figure 8(b), houses the mounting of DC-DC converters and front end electronics, processing electronic and S/C interface electronics.

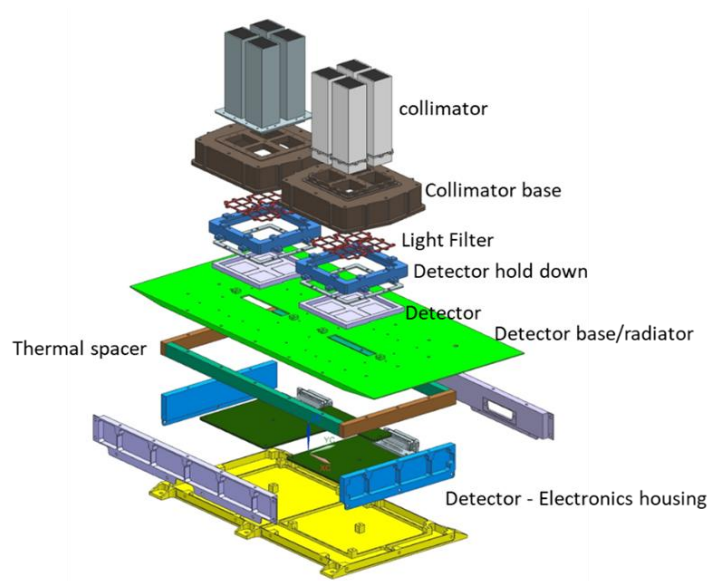

*Figure 7: Detector configuration - exploded model showing all the elements.*

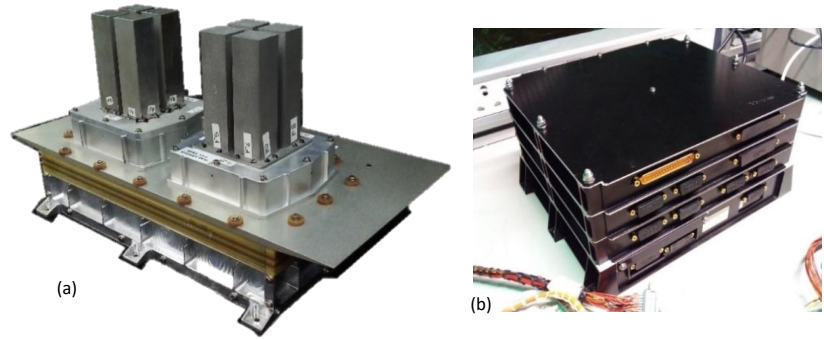

*Figure 8: Realised hardware; (a) Detector package (two such units make XSPECT instrument) and (b) electronics package.*

*2.7 Thermal design*

The two quad modules of each detector package dissipate a maximum of 2 W power. The quad modules are directly clamped on an aluminium plate, which acts like a radiator. Total radiator area of about 800 cm$^2$ for each detector package. The radiator is exposed to deep space during the source observation mode. The area is optimised to achieve the desired nominal operating temperature (< -20°C) throughout the source observation time

## 3. Particle background and radiation damage

Swept charge devices (SCDs) are similar in structure to X-ray CCDs and are prone to radiation damage in space. Particle radiation encountered by the spacecraft from GCR and higher energy protons during SAA transit is expected to affect the SCD performance in the long term. Spenvis analysis [13] was carried out for an orbit of 650 km altitude and six-degree inclination to estimate the incident radiation spectra and the radiation damage. The dose is estimated for shielding of different equivalent aluminium thicknesses using [14]. Total Ionising Dose (TID) is estimated to be within 1 krad, which is benign. 10 MeV equivalent proton fluence that can induce displacement damage in the detector is estimated to be ~1.5E8/cm2 for 2 years of onboard operation.

From the literature and research papers published by Open University [15], UK for damage caused by proton fluence of 3E8/cm2, the FWHM energy resolution degrades to 320 eV at 8 keV. With this we estimate a FWHM resolution of ~250 eV at 8 keV at the end of two years operation of XSPECT. These estimates are at -20° C. And the degradation in FWHM shall be lesser at the expected operating temperatures, which are much below -20° C onboard. Similar results have been reported by HXMT team [16].

## 4. Ground testing and calibration

During test and evaluation phase of the instrument development, environmental tests to space qualification levels and performance checks were carried out. Detailed tests and calibration were carried out at various temperatures during thermo – vacuum tests. Both the detector packages and the electronics package were subjected to thermo-vacuum cycling as per the space qualification requirement.

The schematic of the test arrangement is shown in figure 9. An X-ray gun (Amptek mini-X X-ray source) mounted on a side port of the vaccum chamber was used to generate X-ray Fluorescence (XRF) spectrum from a brass sheet kept at ~ 45 degrees in front of the detector. Flourescence X-rays from the brass sheet illuminated all the 16 SCDs simultaneously. Spectra were measured as a function of device temperature as well as electronics box temperature.

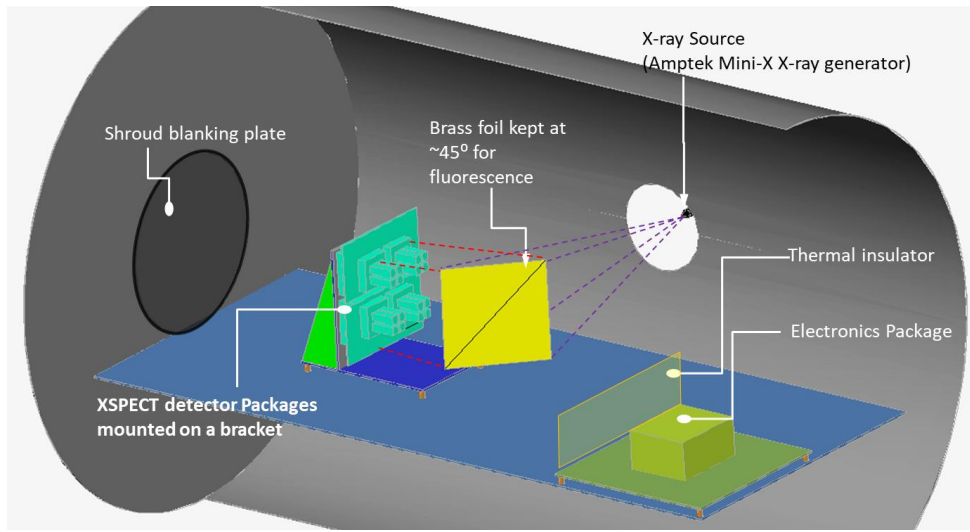

*Figure 9: A schematic of the experimental arrangements during instrument thermos-vacuum tests.*

During the calibration studies first the electronics package base temperature was set to a constant value and then the radiator/detector temperature was increased or decreased. The heating/cooling rate was deliberately made slow, about eight degrees/hour, while data was continuously recorded. The following parameters were derived.

1. Gain (energy-channel relationship) of each device as a function of temperature determined. The gain is to be determined using standard known X-ray energies for environmental conditions which the payload is expected to operate. This constant (for every device) has a unit of eV/channel after digitization of the signal. The gain constant for all devices (except the one deliberately blocked with tantalum sheet) determined by illumination with X-rays from brass foil. From the X-ray events of known energies the channels are mapped into energy space. This relation is sensitive to the temperature of the device as well electronics box temperature.
2. The measure of energy resolution i.e., Full Width at Half Maximum (FWHM) as a function of energy and temperature for all SCDs are determined from the known energy photo-peaks. As temperature of the device increases, the FWHM broadens.

The above two studies were carried out for different electronics and device temperatures shown in the table 2 below.

*Table 2: Temperature values at which device test and calibration carried out*

| Electronics package base temperature | Detector Temperature |
| --- | --- |
| 0⁰C | -15⁰C to -55⁰C |
| 15⁰C | -15⁰C to -55⁰C |
| 30⁰C | -15⁰C to -55⁰C |
| 40⁰C | -15⁰C to -55⁰C |

Copper and zinc are major constituents of Brass and their transition X-ray energies; 8041eV (Cu-K$\alpha$), 8631eV (Zn-K$\alpha$), 8905eV(Cu-K$\beta$) & 9572eV(Zn-K$\beta$) are used for energy calibration. SCDs are clocked devices, i.e., when there is no X-ray photon interaction, a sample is generated for '0 eV' energy. This forms 'zero peak' in the spectra, which is due to the dark signal generated. This zero peak, which has a finite peak channel value, is also used for energy calibration.

Data was continuously recorded with the X-ray source ON for a set electronics base temperature. Then the data was segregated for device temperature bin of 2⁰ C. 'Zero peak' value for every SCD available in the data packet was subtracted from the digitized samples of X-ray events. From this, corrected spectrum was generated. From this the gain and FWHM were estimated for every device for every detector temperature and package temperature.

## 5. Results

Figure 10(a) shows the recorded flourescence spectrum recorded for one of the devices. Known full energy peaks are fitted with Gaussian to derive FWHM and peak channel. A linear fit to energy - peak channel data, shown in figure 10(b) gives gain value of the device at a given temperature. Thus the gain value is derived for each of the devices for the entire operating temperature range of the device. Similarly, it is derived for different electronics temperatures too, though the change is much smaller here. These are shown in figures 11(a) & (b) respectively. The derived gain value matrix at various temperatures for each device is used for analysing the on-board spectral data.

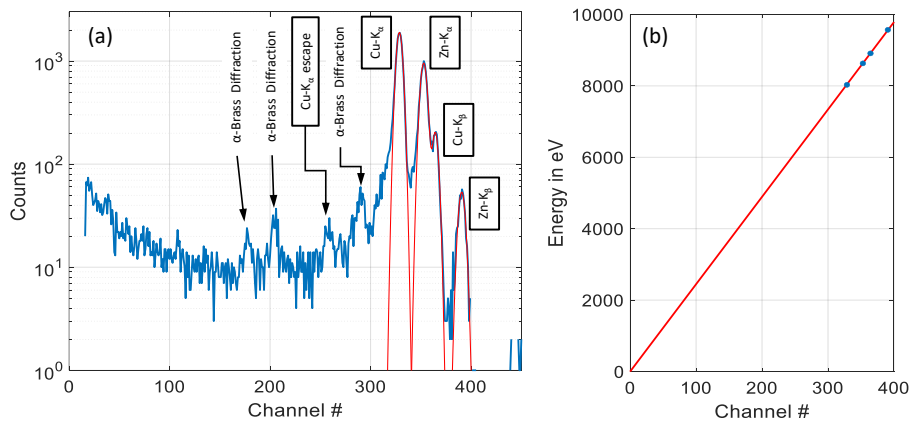

*Figure 10: (a) A sample spectrum recorded for one of the devices (SCD 07) at Electronics Package base temperature of 15° C; measured device temperature was -30° C. Known energy peaks of Cu and Zn fluorescence are used for energy channel calibration. There are lines at lower energies corresponding to an escape peak and diffraction peaks from the alpha brass sheet. (b) The linear fit to known energy and the corresponding peak channel.*

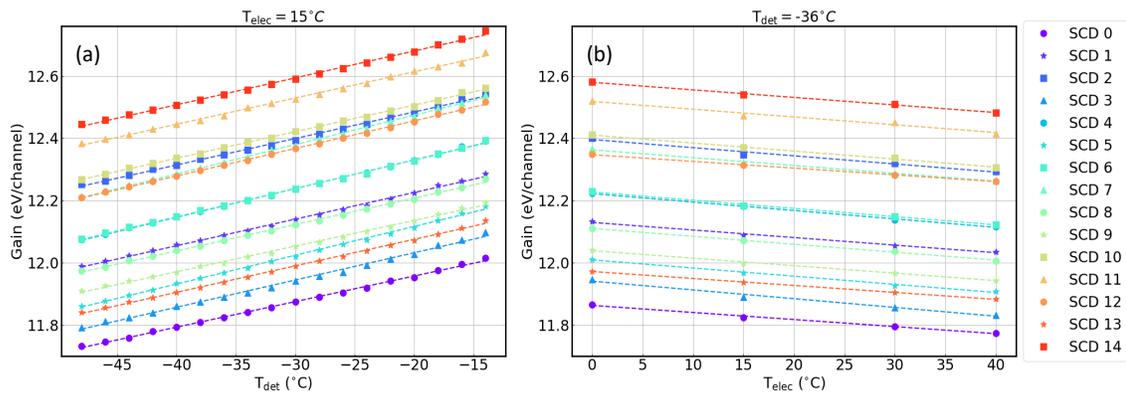

*Figure 11: FWHM in eV for each of the devices at different device temperatures for Copper K-alpha.*

The FWHM derived from the spectra of all the sixteen SCDs at different device temperatures is shown in figure 12. While device to device variation is seen the value is within the science requirement, i.e., ~ 200 eV at nominal operating temperature of -20º C, which corresponds to ~216 eV at 8.05 keV (for Cu-Kα used in these studies).

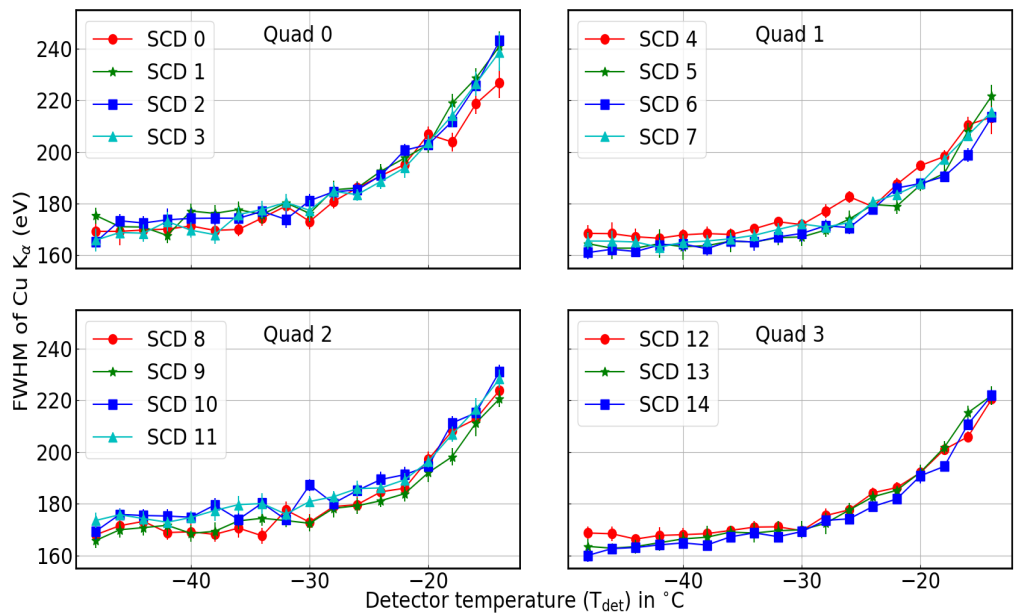

*Figure 12: FWHM in eV for each of the devices at different device temperatures for Copper K-alpha.*

## 6. On-board performance – first results

Some bright cosmic X-ray sources, which includes NS-LMXBs, NS-HMXBs, blackhole X-ray binaries have been identified and observed with XSPECT and POLIX. X-ray intensity variations over 2-4 weeks, period search, power-spectral studies, spectral modelling and its evolution is being performed for these bright X-ray sources. The first X-ray source observed with the XSPECT was super-nova remnant (SNR) Cassiopeia A (also known as Cas A) that shows different emission lines, corresponding to elements such as Magnesium (Mg), Silicon (Si), Sulphur (S), Argon (Ar), Calcium (Ca), and Iron (Fe) as major lines. This is used to verify the gain and FWHM calibration. Subsequently, Tycho SNR and Crab-pulsar were observed during the performance verification phase. The first light (Cas A spectrum) obtained with XSPECT is shown in figure 13(a). In Figure 13(b) the Crab spectrum observed with XSPECT instrument is shown. Cas A spectrum is fitted with a bremsstrahlung to model the continuum and twelve Gaussian components for the emission lines. After applying the ground calibration and fitting the observed spectra, a good match has been found for the line energies and widths. Crab spectrum is fitted with a canonical absorbed power law model. In both cases, good fits are obtained with no major systematics in the residuals. XSPECT has also observed several blank sky regions for estimation of the on-board background. From these observations, the estimated 5σ Sensitivity is 0.6 mCrab for an exposure time of 10 ks.

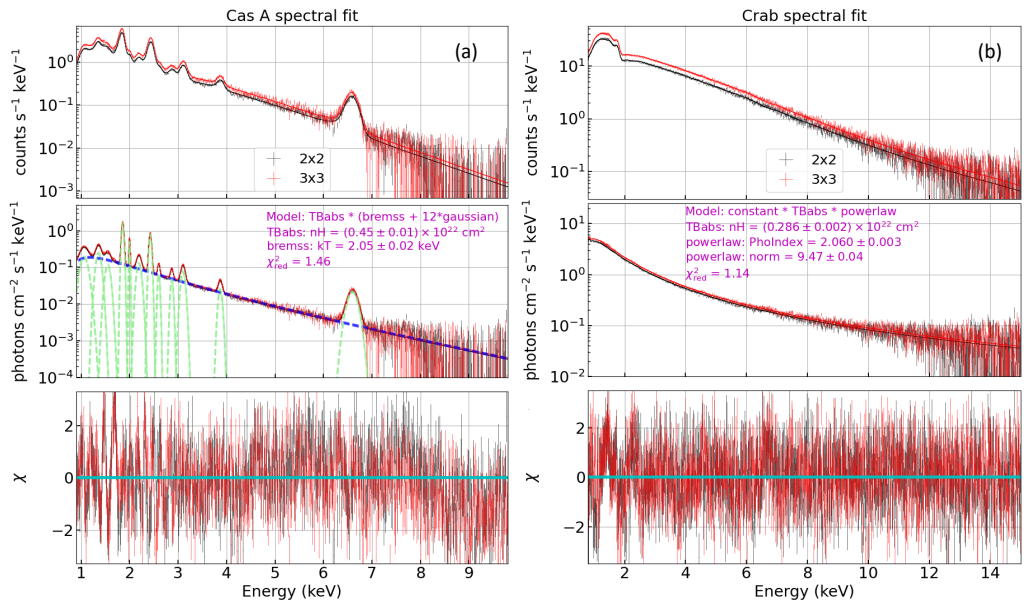

Figure 13: (a) First light source observed by XSPECT is Cas A. (b) Crab spectrum observed by XSPECT instrument

The onboard spectra obtained demonstrate the instrument capability and potential for science observation. Detailed calibration and science outcomes are under preparation and will be communicated separately.

## 7. XSPECT data products

The raw (Level 0) data generated by XSPECT is received at the Payload Operation Center (POC) at Space Astronomy Group, URSC and processed to generate higher level data products. Two levels of processing are defined. Level – 1 (L1) data which is Calibrated data products and auxiliary files include Calibrated, time-tagged event files, Housekeeping data, Attitude file, Orbit file, Filter file and Nominal GTI (good time intervals) file. Level-2 (L2) is scientific data products consists of Filtered event list, Spectra of the two FOVs, and Light curve of the two FOVs.

Nominally, the target being observed by XSPECT does not change within a day. Hence, both L1 and L2 products are generated day-wise. In case of multiple source observations, however, the products will also be segregated by the target. All the data products are in FITS format, and compatible with the standard high-energy data analysis software HEASoft [https://heasarc.gsfc.nasa.gov/docs/software/lheasoft/]. The L1 and L2 products generated at the POC will be sent to the Indian Space Science Data Center (ISSDC) for archival and dissemination.

## 8. Conclusions

XSPECT instrument on XPoSat is a unique opportunity to observe bright sources for longer duration to evaluate spectral variability. The payload is configured with SCDs providing larger area with passive cooling but providing good spectral resolution. Two different FOV collimators are used to derive better local source background. Detailed ground studies have been carried out to demonstrate instrument performance and for ground calibration. The results from initial source observations prove the potential of the instrument for further science observations. The science data as well as processing software will be made available for the scientific community.

*Disclosures*

The authors declare that there are no financial interests, commercial affiliations, or other potential conflicts of interest that could have influenced the objectivity of this research or the writing of this paper.

*Code, Data, and Materials Availability*

The data utilized in this study from ground experiments and calibration are available from the authors upon request. The source observation data used are also presently available from authors upon request while they are being made available from https://pradan.issdc.gov.in/pradan/.

*Acknowledgements:*

We thank XPoSat project team, assembly, integration and checkout teams, mission team and the facilities team for their involvement and support in enabling XSPECT payload on XPoSat mission. Authors greately acknowledge the support from the project directors Sri Amareshwar Khened and Sri Brindaban Mahto. We thank Director, U R Rao Satellite Centre, Deputy Director, PDMSA, and Group Head, SAG for their reviews and support.